\documentclass[12pt]{article}
\usepackage{epsfig}
\def\bb {\begin {eqnarray}}
\def\ee {\end {eqnarray}}

\newcommand{\HS}{S}
\newcommand{\ep}{\varepsilon}
\newcommand{\bea}{\begin{eqnarray}}
\newcommand{\eea}{\end{eqnarray}}
\newcommand{\z}{&&\hspace*{-1cm}}
\begin{document}

\begin{center}
{\Large {\bf The property of maximal transcendentality
in the  ${\mathcal N}=4$
Supersymmetric Yang-Mills
}}
\\ \vspace*{5mm} A.~V.~Kotikov
\end{center}

\begin{center}
Bogoliubov Laboratory of Theoretical Physics \\
Joint Institute for Nuclear Research\\
141980 Dubna, Russia
\end{center}

\begin{abstract}
We present results for the universal anomalous dimension
$\gamma _{uni}(j)\,$ of Wilson twist-2 operators in the
${\mathcal N}=4$ Supersymmetric Yang-Mills theory in the first four
orders of perturbation theory.
\end{abstract}


\section{Introduction}

This paper deals with the study of the properties of the 
 Balitsky-Fadin-Kuraev-Lipatov (BFKL) \cite{BFKL}
and Dokshitzer-Gribov-Lipatov-Altarelli-Parisi (DGLAP) \cite{DGLAP}
equations in 
the  ${\mathcal N}=4$
Supersymmetric Yang-Mills (SYM) model \cite{BSSGSO}.

Lev Nikolaevich Lipatov made a fundamental
contributions in discovery of 
both the equations 
as in the framework of Quantum
Chromodynamics (QCD)
and later to its supersymmetric extensions.

The BFKL and DGLAP equations resum, respectively, the most important 
contributions
$\sim \alpha_s \ln(1/x)$ and $\sim \alpha_s \ln(Q^2/\Lambda^2)$ in different
kinematical regions of the Bjorken variable $x$ and the ``mass'' $Q^2$ of the
virtual photon in the 
deep inelastic scattering (DIS) 
and, thus, they are the cornerstone in analyses of
the experimental data
from lepton-nucleon and nucleon-nucleon scattering processes.

In the supersymmetric cases the equations are simplified drastically
and in the
${\mathcal N}=4$ SYM they
become to be related each others for the nonphysical values of 
Mellin moments 
$j$ as it has been proposed 
Lipatov in \cite{KL}.

The purpose of this paper is 
to show the results for the anomalous dimension matrix
of the twist-2 Wilson operators, which have been
obtained by author in collaboration with
Lev Nikolaevich Lipatov during last 10 years.

The anomalous dimensions 
govern the
Bjorken scaling violation for parton distributions in a framework of 
QCD. These quantities are
given by the Mellin transformation (the symbol $\tilde{} $ is used for
spin-dependent case and $a_s=\alpha_s/(4\pi)$)
\begin{eqnarray}
\gamma _{ab}(j) &=& \int_{0}^{1}dx\,\,x^{j-1} W_{b\rightarrow a}(x)
~=~ \gamma^{(0)} _{ab}(j) a_s + \gamma^{(1)} _{ab}(j) a_s^2+
\gamma^{(2)} _{ab}(j) a_s^3 + O(a_s^4),~~
\nonumber \\
\tilde{\gamma} _{ab}(j) &=& \int_{0}^{1}dx\,\,x^{j-1}
\tilde{W}_{b\rightarrow a}(x)
~=~ \tilde{\gamma}^{(0)} _{ab}(j) a_s + \tilde{\gamma}^{(1)} _{ab}(j) a_s^2+
\tilde{\gamma}^{(2)} _{ab}(j) a_s^3 + O(a_s^4)
\end{eqnarray}
of the splitting kernels $W_{b\rightarrow a}(x)$ and
$\tilde{W}_{b\rightarrow a}(x)$ for the DGLAP
equation~\cite{DGLAP}
which evolves the parton densities $f_{a}(x,Q^{2})$ and
$\tilde{f}_{a}(x,Q^{2})$
(hereafter $a=\lambda,\,g,\,\phi$ for
the spinor, vector and scalar particles, respectively
\footnote{In the spin-dependent case $a=\lambda,\,g$.})
as follows
\begin{eqnarray}
\frac{d}{d\ln {Q^{2}}}f_{a}(x,Q^{2}) &=& \int_{x}^{1}\frac{dy}{y}
\sum_{b}W_{b\rightarrow a}(x/y)\,f_{b}(y,Q^{2}) \, ,
\nonumber \\
\frac{d}{d\ln {Q^{2}}}\tilde{f}_{a}(x,Q^{2}) &=& \int_{x}^{1}\frac{dy}{y}
\sum_{b}\tilde{W}_{b\rightarrow a}(x/y)\,\tilde{f}_{b}(y,Q^{2}) \,.
\label{DGLAP}
\end{eqnarray}
The anomalous dimensions and splitting kernels in QCD are known up to the
next-to-next-to-leading order (NNLO) of the perturbation theory
(see~\cite{VMV} and references therein).

The QCD expressions for anomalous dimensions can be transformed to the case
of the
${\mathcal N}$-extended Supersymmetric Yang-Mills theories (SYM)
if one will use for the Casimir operators $C_{A},C_{F},T_{f}$ the following
values $C_{A}=C_{F}=N_{c}$, $T_{f}n_f={\mathcal N}N_{c}/2$.
For ${\mathcal N}\!\!=\!\!2$ and ${\mathcal N}\!\!=\!\!4$-extended SYM the anomalous dimensions of the Wilson operators get also
additional contributions coming from scalar particles~\cite{KL}.
These anomalous dimensions were calculated in the next-to-leading order
(NLO)~\cite{KL,KoLiVe}
for the ${\mathcal N}=4$ SYM.

However, it turns out, that the expressions
for eigenvalues of the anomalous dimension matrix in the ${\mathcal N}=4$ SYM 
\cite{BSSGSO} can be derived directly
from the QCD anomalous dimensions without tedious calculations by using a
number of plausible arguments. The method elaborated in Ref.~\cite{KL} for
this purpose is based on special properties of
the integral kernel for
the BFKL
equation~\cite{BFKL,next,KL00}
 in this model
and a new relation between the BFKL and DGLAP equations (see~\cite{KL}).
In the NLO approximation this method gives the correct results for
anomalous dimensions eigenvalues, which was checked by {\it direct calculations} in Ref.~\cite{KoLiVe}.
Using the results for the NNLO
corrections to anomalous dimensions in QCD~\cite{VMV}
and the method of Ref.~\cite{KL} we derive the
eigenvalues of the anomalous dimension matrix for the ${\mathcal N}=4$ SYM in the NNLO
approximation~\cite{KLOV}.

Starting from four loops, i.e. above existing QCD calculations,
the corresponding results for the anomalous dimensions
 can be obtained (see \cite{KLRSV,BaJaLu,LuReVe}) from
the long-range asymptotic Bethe equations
together with some additional terms, so-called {\it wrapping
corrections}, coming in agreement with Luscher approach.
\footnote{The three- and four-loop results for the universal  anomalous
dimension have been reproduced (see \cite{KoReZi}) also by solution of
so-called Baxter equation, which can be obtained from
the long-range asymptotic Bethe equations.}

The obtained result is very important for the verification of the
various assumptions (see recent reviews \cite{Staudacher:2004tk}--\cite{Rej}
and references therein)
coming from the investigations
of the properties of a conformal operators in the context of AdS/CFT
correspondence~\cite{AdS-CFT}.

The paper is organized as follows.
In Section 2 we discuss the BFKL equation, the leading order anomalous 
dimensions of Wilson operators and propose the method of obtaining the 
eigenvalues of the anomalous dimension matrix above the leading order.
Section 3 contains the calculations of some Feynman diagrams by similar 
method. In Section 4 we consider three-loop results for the universal
anomalous dimension taking from the corresponding calculations in QCD.
Four-loop corrections to  the universal
anomalous dimension are considered in Section 5.

\section{Evolution equation in ${\mathcal N}=4$ SYM}

The reason to investigate the BFKL and DGLAP equations in the case of
supersymmetric theories is
related to
a common belief, that the high symmetry
may significantly simplify their structure. Indeed, it was
found in the leading order (LO)
approximation
~\cite{BFKL2}, that the
so-called quasi-partonic operators in ${\mathcal N}=1$ SYM are unified in
supermultiplets with anomalous dimensions obtained from 
some universal
anomalous dimension 
by shifting its argument by an
integer number. Further, the anomalous dimension matrices for twist-2
operators are fixed by the superconformal invariance~\cite{BFKL2}.
Calculations in the maximally extended ${\mathcal N}=4$ SYM, where the
coupling constant is not renormalized, give even more remarkable
results. Namely, it turns out, that here all twist-2 operators enter in
the same multiplet, their anomalous dimension matrix is fixed completely
by the super-conformal invariance  and its universal anomalous dimension
in LO is proportional to $\Psi (j-1)-\Psi (1)$ (see the
subsection 2.2),
which means, that the
evolution equations for the matrix elements of quasi-partonic operators in
the multicolour limit $N_{c}\rightarrow \infty $ are equivalent to
the Schr\"{o}dinger equation for
an integrable Heisenberg spin model~\cite{N=4,LN4}. In QCD the
integrability remains only in a small sector of these operators~\cite{BDMB} (see also~\cite{Ferretti:2004ba}). In the case of ${\mathcal N}=4$ SYM
the equations for other sets of operators are also
integrable~\cite{Minahan:2002ve,Beisert:2003tq}.

Similar results related
to the integrability of the multi-colour QCD were obtained
earlier in the Regge limit~\cite{Integr}. Moreover, it was shown~\cite{KL},
that in the ${\mathcal N}=4$ SYM there is a deep relation between the BFKL 
and DGLAP evolution equations. Namely, the $j$-plane singularities of 
anomalous dimensions of the Wilson
twist-2 operators in this case can be obtained from the eigenvalues of the
BFKL kernel by their analytic continuation. The NLO
calculations in ${\mathcal N}=4$ SYM demonstrated~\cite{KL}, that some of these
relations are valid also in higher orders of perturbation theory. In
particular, the BFKL equation has the property of the hermitian
separability, the linear combinations of the multiplicatively renormalized
operators do not depend on the coupling constant, the eigenvalues of the
anomalous dimension matrix are expressed in terms of the universal
function $\gamma _{uni}(j)$ which can be obtained also from the BFKL
equation~\cite{KL}.

\subsection{BFKL
}

To begin with, we review shortly the results of Refs. \cite{next,KL00},
where the QCD
radiative corrections to the BFKL integral kernel at $t=0$ were calculated.
\footnote{The $t\neq 0$ case can be found in the recent papers 
\cite{Fadin:2007xy}.}
We discuss only the formulae important for our analysis.

The total cross-section $\sigma (s)$ for the high energy scattering of
colourless particles $A,B$ written in terms of their impact factors $\Phi
_{i}(q_{i})$ and the $t$-channel partial wave $G_{\omega }(q,q^{\prime })$
for the gluon-gluon scattering is
\begin{equation}
\sigma (s)~=~\int \frac{d^{2}q\,d^{2}q^{\prime }}{(2\pi
)^{2}\,q^{2}\,q^{\prime 2}}\Phi _{A}(q)\,\Phi _{B}(q^{\prime
})\int_{a-i\infty }^{a+i\infty }\frac{d\omega }{2\pi i}\left(
{\frac{s}{s_0}}\right) ^{\omega }G_{\omega }(q,q^{\prime }), ~~~
s_0 = |q||q^{\prime }|.
\label{BFKL}
\end{equation}

Here $q$ and $q^{\prime }$ are transverse momenta\footnote{%
To simplify equations hereafter
we omit arrows
in the notation of transverse momenta $\overrightarrow{q},~ \overrightarrow{%
q^{\prime}},~ \overrightarrow{q_1},~\overrightarrow{q_2},~ ...~$, i.e. in
our formulae the momenta $\overrightarrow{q},~
\overrightarrow{q^{\prime}},~%
\overrightarrow{ q_1},~\overrightarrow{q_2},~ ...~$ will be represented as
$q,~q^{\prime},~q_1,~q_2,~...~$, respectively. Note, however, that the
momenta $p_A$ and $p_B$ are $D$-space momenta.} of virtual gluons and
$s=2p_{A}p_{B}$ is the squared invariant mass for the colliding particle
momenta $p_{A}$ and $p_{B}$.

Using the dimensional regularization in the $\overline{MS}$-scheme to
 remove ultraviolet and infrared divergences in
intermediate expressions, the BFKL equation for $G_{\omega }(q,q^{\prime })$
can be written in the following form
\begin{equation}
\omega G_{\omega }(q,q_{1})~=~\delta ^{D-2}(q-q_{1})+\int
d^{D-2}q_{2}\,K(q,q_{2})\,G_{\omega }(q_{2},q_{1})\,,
\end{equation}
where
\begin{equation}
K(q_{1},q_{2})~=~2\,\omega (q_{1})\,\delta
^{D-2}(q_{1}-q_{2})+K_{r}(q_{1},q_{2})
\end{equation}
and the space-time dimension $D=4-2\varepsilon $ for $\varepsilon \to 0$.
The gluon Regge trajectory $\omega (q)$ and the integral kernel $%
K_{r}(q_{1},q_{2})$ related to the real particle production have been
calculated in 
\cite{LF89}-\cite{CCF}.

As it was shown in \cite{next,KL00}, a complete and orthogonal set of
eigenfunctions of the homogeneous BFKL equation in 
LO is
\begin{equation}
G_{n,\gamma }(q/q^{\prime },\theta )~=~\left( \frac{q^{2}}{q^{\prime 2}}
\right) ^{\gamma -1}e^{in\theta }
\end{equation}

The BFKL kernel in this representation is diagonalized up to the effects
related with the running coupling constant $a_{s}(q^{2})$:
\bea
\omega^{QCD}_{\overline{MS}} = 4 a_{s}(q^{2})\biggl[
\chi (n,\gamma )+\delta^{QCD}_{\overline{MS}} (n,\gamma )
a_{s}(q^{2}) \biggr] \,.
\label{ome1a}
\eea

Applying formulae of \cite{KL00},
we obtain the following results for eigenvalues 
(23):
\begin{eqnarray}
\chi (n,\gamma ) &=&2\Psi (1)-\Psi \Bigl(\gamma +\frac{n}{2}\Bigr)-\Psi
\Bigl%
(1-\gamma +\frac{n}{2}\Bigr)  \label{7} \\
&&  \nonumber \\
\delta^{QCD}_{\overline{MS}} (n,\gamma ) &=&
\biggl(\frac{67}{9} -2\zeta (2)
-\frac{10}{9} \,
\frac{n_{f}}{N_{c}}
\biggr) %
\chi (n,\gamma )
+6\zeta (3)+
 \Psi ^{\prime \prime }\Bigl(\gamma +\frac{n}{2}\Bigr)+ \Psi ^{\prime
\prime } \Bigl(1-\gamma +\frac{n}{2}\biggr)
  \nonumber \\
&-&  2\Phi (n,\gamma )-2\Phi (n,1-\gamma )
-\biggl(\frac{11}{3}-
\frac{2}{3} \,
\frac{n_{f}}{N_{c}}
\biggr) \frac{1}{2}
\chi ^{2}(n,\gamma )
\nonumber \\
&+&\frac{\pi ^{2}\cos (\pi \gamma )}{\sin ^{2}(\pi \gamma
)(1-2\gamma )}\Biggl\{
\biggl(1
+\frac{\tilde{n}_{f}}{N_{c}^{3}}
\biggr) \frac{\gamma (1-\gamma )}{
2(3-2\gamma )(1+2\gamma )}\cdot \delta _{n}^{2}
  \nonumber \\
&-& \biggl(3+\biggl(1
+\frac{\tilde{n}_{f}}{N_{c}^{3}}
\biggr)
\frac{2+3\gamma (1-\gamma )}{(3-2\gamma )(1+2\gamma )}
\biggr)\cdot \delta_{n}^{0}
\Biggr\},
  \label{8.2}
\end{eqnarray}
where
$\delta _{n}^{m}$ is the Kroneker symbol, and $\Psi (z)$, $\Psi
^{\prime }(z)$ and $\Psi ^{\prime \prime }(z)$ are the Euler $\Psi $
-function and its  derivatives. The function $\Phi (n,\gamma
)$ is given below
\begin{eqnarray}
\z \Phi (n,\gamma ) ~=~
~\sum_{k=0}^{\infty }\frac{(-1)^{k+1}}{%
k+\gamma +n/2}\Biggl[ \Psi ^{\prime }(k+n+1)-\Psi ^{\prime }(k+1)
\nonumber \\
\z
+(-1)^{k}
\Bigl(\beta ^{\prime }(k+n+1)+\beta ^{\prime }(k+1)\Bigr)\biggr)
- \frac{1}{k+\gamma +n/2}\biggl( \Psi (k+n+1)-\Psi (k+1)%
\biggr) \Biggr]  \label{9}
\end{eqnarray}
and
\[
\beta ^{\prime }(z)=\frac{1}{4}\Biggl[ \Psi ^{\prime }\Bigl(\frac{z+1}{2}%
\Bigr)-\Psi ^{\prime }\Bigl(\frac{z}{2}\Bigr)\Biggr]
\]

Adding contributions of scalars and transforming 
fermions
from fundamental to adjoint representation, we can obtain the BFKL form
(\ref{ome1a}) in ${\mathcal N}=4$ SYM in 
$\overline{DR}$ scheme \cite{DRED}
\begin{eqnarray}
 \delta ^{N=4}_{\overline{DR}}(n,\gamma )&=&
6\zeta (3)+
 \Psi ^{\prime \prime }\Bigl(\gamma +\frac{n}{2}\Bigr)+ \Psi ^{\prime
\prime } \Bigl(1-\gamma +\frac{n}{2}\biggr)
  \nonumber \\
&-&  2\Phi (n,\gamma )-2\Phi (n,1-\gamma ) -
2\zeta (2) \chi (n,\gamma ),   \label{K3}
\end{eqnarray}
where the $\overline{DR}$
coupling constant $\hat{a}_{s}$ is related \cite{Altarelli} with
the $\overline{MS}$ one  $a_{s}$ as
\begin{eqnarray}
\hat{a}_{s} ~=~ a_{s} + \frac{1}{3} a^2_{s}.
\label{9aa}
\end{eqnarray}

Note that the sum $\Phi (n,\gamma )+\Phi (n,1-\gamma )$
 can be rewritten (see \cite{KL}) as a combination of functions with
argument dependent on $\gamma +n/2 \equiv M$ and
$1-\gamma +n/2 \equiv \tilde{M}$. Indeed
\begin{eqnarray}
&&\Phi (n,\gamma )+\Phi (n,1-\gamma )
=\chi (n,\gamma )\,\left( \beta
^{\prime }(M)+\beta ^{\prime }(1-\widetilde{M})\right)   \nonumber \\
&&+\Phi _{2}(M)-\beta ^{\prime }(M)\left[ \Psi (1)-\Psi (M)\right] +\Phi
_{2}(1-\widetilde{M})-\beta ^{\prime }(1-\widetilde{M})\left[ \Psi (1)-\Psi (1-%
\widetilde{M})\right] ,  \nonumber
\end{eqnarray}
where $\chi (n,\gamma )$ is given by Eq.(\ref{7}) and
\begin{eqnarray}
\Phi_2 (M )=~\sum_{k=0}^\infty
\frac{\left( \beta ^{\prime }(k+1)+
(-1)^k \Psi ^{\prime }(k+1)\right) }{k+M}
-\sum_{k=0}^\infty \frac{(-1)^k\left( \Psi (k+1)-\Psi (1)\right) }{%
(k+M)^2} \,,
  \label{9.1}
\end{eqnarray}

So, this transformation leads to the hermitian separability of BFKL equation
in ${\mathcal N}=4$ SYM (see Ref. \cite{KL} and discussions therein).


\subsection{Leading order anomalous dimension matrix in ${\mathcal N}=4$ SYM}

In the ${\mathcal N}=4$ SYM theory~\cite{BSSGSO}
one can introduce the following colour and $SU(4)$ singlet local Wilson twist-2
operators~\cite{KL,KoLiVe}:
\begin{eqnarray}
\mathcal{O}_{\mu _{1},...,\mu _{j}}^{g} &=&\hat{S}
G_{\rho \mu_{1}}^{a}{\mathcal D}_{\mu _{2}}
{\mathcal D}_{\mu _{3}}...{\mathcal D}_{\mu _{j-1}}G_{\rho \mu _{j}}^a\,,
\label{ggs}\\
{\tilde{\mathcal{O}}}_{\mu _{1},...,\mu _{j}}^{g} &=&\hat{S}
G_{\rho \mu_{1}}^a {\mathcal D}_{\mu _{2}}
{\mathcal D}_{\mu _{3}}...{\mathcal D}_{\mu _{j-1}}{\tilde{G}}_{\rho \mu _{j}}^a\,,
\label{ggp}\\
\mathcal{O}_{\mu _{1},...,\mu _{j}}^{\lambda } &=&\hat{S}
\bar{\lambda}_{i}^{a}\gamma _{\mu _{1}}
{\mathcal D}_{\mu _{2}}...{\mathcal D}_{\mu _{j}}\lambda ^{a\;i}\,, \label{qqs}\\
{\tilde{\mathcal{O}}}_{\mu _{1},...,\mu _{j}}^{\lambda } &=&\hat{S}
\bar{\lambda}_{i}^{a}\gamma _{5}\gamma _{\mu _{1}}{\mathcal D}_{\mu _{2}}...
{\mathcal D}_{\mu_{j}}\lambda ^{a\;i}\,, \label{qqp}\\
\mathcal{O}_{\mu _{1},...,\mu _{j}}^{\phi } &=&\hat{S}
\bar{\phi}_{r}^{a}{\mathcal D}_{\mu _{1}}
{\mathcal D}_{\mu _{2}}...{\mathcal D}_{\mu _{j}}\phi _{r}^{a}\,,\label{phphs}
\end{eqnarray}
where ${\mathcal D}_{\mu }$ are covariant derivatives.
The spinors $\lambda _{i}$ and
field tensor $G_{\rho \mu }$ describe gluinos and gluons, respectively, and
$\phi _{r}$ are the complex scalar fields.
For all operators in Eqs.~(\ref{ggs})-(\ref{phphs}) the symmetrization of the tensors
in the Lorentz indices
$\mu_{1},...,\mu _{j}$ and a subtraction of their traces is assumed.

The elements of the LO anomalous dimension matrix in the ${\mathcal N}=4$
SYM have
the following form (see \cite{LN4}):

for tensor twist-2 operators
\begin{eqnarray}
\gamma^{(0)}_{gg}(j) &=& 4
\left( \Psi(1)-\Psi(j-1)-\frac{2}{j}+\frac{1}{j+1}
-\frac{1}{j+2} \right),  \nonumber \\
\gamma^{(0)}_{\lambda g}(j) &=& 8 \left(\frac{1}{j}-\frac{2}{j+1}+\frac{2}{j+2}
\right),~~~~~~~~~~\, \gamma^{(0)}_{\varphi g}(j) ~=~ 12 \left( \frac{1}{j+1}
-\frac{1}{j+2} \right),  \nonumber \\
\gamma^{(0)}_{g\lambda}(j) &=& 2 \left(\frac{2}{j-1}-\frac{2}{j}+\frac{1}{j+1}
\right),~~~~~~~~~~\, \gamma^{(0)}_{q\varphi}(j) ~=~ \frac{8}{j} \,,
\nonumber \\
\gamma^{(0)}_{\lambda \lambda}(j) &=& 4 \left( \Psi(1)-\Psi(j)+\frac{1}{j}-
\frac{2}{j+1}%
\right),~~ \gamma^{(0)}_{\varphi \lambda}(j) ~=~ \frac{6}{j+1} \,, \nonumber \\
\gamma^{(0)}_{\varphi \varphi}(j) &=& 4 \left( \Psi(1)-\Psi(j+1)\right),
~~~~~~~~~~~~~~~\, \gamma^{(0)}_{g\varphi}(j) ~=~ 4
\left(\frac{1}{j-1}-\frac{%
1}{j} \right),  \label{3.2}
\end{eqnarray}

for the pseudo-tensor operators:
\begin{eqnarray}
\widetilde \gamma^{(0)}_{gg}(j) &=& 4 \left( \Psi(1)-\Psi(j+1)-\frac{2}{j+1} +%
\frac{2}{j} \right),  \nonumber \\
\widetilde \gamma^{a,(0)}_{\lambda g}(j) &=&
8 \left(-\frac{1}{j}+\frac{2}{j+1}\right),
~~~~~ \widetilde \gamma^{(0)}_{g\lambda}(j) ~=~
2\left( \frac{2}{j} -\frac{1}{j+1}
\right),  \nonumber \\
\widetilde \gamma^{(0)}_{\lambda \lambda}(j) &=&
4 \left( \Psi(1)-\Psi(j+1)+\frac{1}{j+1}-\frac{1}{j}\right).  \label{3.3}
\end{eqnarray}

The matrices, based on the anomalous dimensions (\ref{3.2}) and  (\ref{3.3}),
can be diagonalized \cite{LN4,KL}. They have the following remarkable form
\begin{eqnarray}
{\Biggl[D\Gamma D^{-1}\Biggr]}^{N=4}_{\mathbf{unpol}} =
\begin{array}{|ccc|}
-4S_1(j-2) & 0 & 0 \\
0  & -4S_1(j) & 0  \\
0 & 0 & -4S_1(j+2)
\end{array}
\nonumber
\end{eqnarray}
\begin{eqnarray}
{\Biggl[D\Gamma D^{-1}\Biggr]}^{N=4}_{\mathbf{pol}} =
\begin{array}{|cc|}
-4S_1(j-1) & 0  \\
0  & -4S_1(j+1)
\end{array}
\,,\nonumber
\end{eqnarray}
where $S_1(j)$ is defined below in (\ref{FI2}).

Thus, the LO anomalous dimensions
of all multiplicatively renormalized operators
can be extracted through one universal function
$$
\gamma^{(0)}_{uni}(j)~=~-4S(j-2) \equiv
-4\Bigl(\Psi(j-1)-\Psi(1) \Bigr) \equiv -4 \sum_{r=1}^{j-2}
\frac{1}{r}. $$

\subsection{Method of obtaining the eigenvalues of the anomalous dimension
matrix in ${\mathcal N}=4$ SYM}\label{MethodAD}

As it was already pointed out in the Introduction, the universal anomalous
dimension can be extracted directly from the QCD results without finding the
scalar particle contribution. This possibility is based on the deep
relation between the DGLAP and BFKL dynamics in the ${\mathcal N}=4$ SYM~
\cite{KL00,KL}.

To begin with, the eigenvalues of the BFKL kernel are
the analytic functions of the conformal spin $\left| n\right| $ 
at least in two first orders of
perturbation theory (see Eqs. (\ref{ome1a}), (\ref{7}) and (\ref{K3})).
Further, in the framework of the ${\overline{\mathrm{DR}}}$-scheme~\cite{DRED}
one can obtain from (\ref{7}) and (\ref{8.2}),
that there is no
mixing among the special functions of different transcendentality levels $i$
\footnote{
Similar arguments were used also in~\cite{FleKoVe} to obtain
analytic results for contributions of some complicated massive Feynman
diagrams without direct calculations (see also the section 3).},
i.e. all special functions at the NLO correction contain only sums of the
terms $\sim 1/\gamma^{i}~(i=3)$. More precisely, if we introduce the
transcendentality level $i$ for the eigenvalues $\omega(\gamma)$ of integral kernels of the BFKL
equations
in an accordance with the complexity of the terms in the
corresponding sums
\[
\Psi \sim 1/\gamma ,~~~\Psi ^{\prime }\sim \beta ^{\prime }\sim \zeta
(2)\sim 1/\gamma ^{2},~~~\Psi ^{\prime \prime }\sim \beta ^{\prime \prime
}\sim \Phi \sim \zeta (3)\sim 1/\gamma ^{3},
\]
then for the BFKL kernel in LO
and in NLO the
corresponding levels are $i=1$ and $i=3$, respectively.

Because in ${\mathcal N}=4$ SYM there is a relation between the BFKL and DGLAP equations
(see~\cite{KL00,KL}), the similar properties should be valid for the
anomalous dimensions themselves, i.e. the basic functions $\gamma
_{uni}^{(0)}(j)$, $\gamma _{uni}^{(1)}(j)$ and $\gamma _{uni}^{(2)}(j)$ are
assumed to be of the types $\sim 1/j^{i}$ with the levels $i=1$, $i=3$ and
$i=5$, respectively. An exception could be for the terms appearing at a given
order from previous orders of the perturbation theory. Such
contributions could be generated and/or removed by an approximate finite
renormalization of the coupling constant. But these terms do not appear in
the ${\overline{\mathrm{DR}}}$-scheme.

It is known, that at the LO and NLO approximations
(with the SUSY relation for the QCD color factors $C_{F}=C_{A}=N_{c}$) the
most complicated contributions (with $i=1$ and $i=3$, respectively) are the
same for all LO and NLO anomalous dimensions in QCD~\cite{VMV}
and for the LO and NLO scalar-scalar anomalous
dimensions~\cite{KoLiVe}. This property allows one to find the
universal anomalous dimensions $\gamma _{uni}^{(0)}(j)$ and $\gamma
_{uni}^{(1)}(j)$ without knowing all elements of the anomalous dimensions
matrix~\cite{KL}, which was verified by the exact calculations in~\cite{KoLiVe}.

Using above arguments, we conclude, that at the NNLO level there is only one
possible candidate for $\gamma _{uni}^{(2)}(j)$. Namely, it is the most
complicated part of the QCD anomalous dimensions matrix
(with
the SUSY relation for the QCD color factors
$C_{F}=C_{A}=N_{c}$).
Indeed, after the diagonalization of the
anomalous dimensions matrix its eigenvalues should have this most complicated part
as a common contribution because they differ each from others only by a shift of
the argument and their differences are constructed from
less complicated terms. The non-diagonal matrix elements of the anomalous dimensions matrix
contain also only less complicated terms (see, for example, anomalous dimensions exact
expressions at LO and NLO approximations
in Refs.~\cite{VMV}
for QCD
and~\cite{KoLiVe} for ${\mathcal N}=4$ SYM) and therefore they cannot generate
the most complicated contributions to the eigenvalues of anomalous dimensions matrix.

Thus, the most complicated part of the NNLO QCD
anomalous dimensions should coincide (up to color factors)
with the universal anomalous dimension $\gamma_{uni}^{(2)}(j)$.

\section{Calculation of Feynman integrals}

Similar arguments
give a possibility to calculate
a large class of Feynman
diagrams,
so-called master-integrals \cite{Broadhurst:1987ei}.
Let us consider it in some details.

Application of the integration-by-part (IBP) procedure \cite{Chetyrkin:1981qh}
to loop internal momenta
leads to
relations between different Feynman integrals (FI) and, thus,
to necessity to calculate only some of them, which in a sense, are
independent (see  \cite{Kotikov:1990zs} ). These independent diagrams
(which were chosen quite arbitrary, of course) are called the
master-integrals \cite{Broadhurst:1987ei}.

The application of the IBP procedure \cite{Chetyrkin:1981qh} to the
master-integrals themselves leads to the differential equations
\cite{Kotikov:1990kg,Kotikov:1991hm} for them with
the inhomogeneous terms (ITs) containing less complicated diagrams.
\footnote{The ``less complicated diagrams'' contain usually less number of
propagators and sometimes they can be represented as diagrams with less
number of loops and with some ``effective masses'' (see, for example, 
\cite{Kniehl:2005bc} and references therein).} The application
of the IBP procedure to these diagrams leads to the new differential
equations for them with the new ITs
containing even farther
less complicated diagrams.
Repeating the procedure several times, at a last step one can obtain the
ITs containing only tadpoles which can be calculated  in-turn very easily.

Solving the
differential equations
at this last step,
one can reproduce the diagrams for ITs of the differential equations at the
previous step.
Repeating the procedure several times one can obtain the results for the
initial Feynman diagram.

This scheme has been used successfully for calculation of two-loop two-point
\cite{Kotikov:1990zs,Fleischer:1997bw}
and three-point diagrams \cite{Fleischer:1997bw,FleKoVe,FleKoVe1}
with one nonzero mass. This procedure is very
powerful but quite complicated. There are, however, some simplifications, which
are based on the series representations of Feynman integrals.

Indeed, the inverse-mass expansion of two-loop two-point and three-point
diagrams
\footnote{We consider only three-point
diagrams with independent momenta $q_1$ and $q_2$, which obey
the conditions
 $q_1^2=q_2^2=0$ and $(q_1+q_2)^2\equiv q^2 \neq 0$.}
with one nonzero mass,
can be considered as
\begin{eqnarray}
\mbox{ FI} ~ && = ~ \frac{\hat{N}}{q^{2\alpha}} \,
\sum_{n=1} \, C_n \, \frac{{(\eta x)}^n}{n^c} \, \biggl\{F_0(n) +
\biggl[ \ln (-x) \, F_{1,1}(n)  +
\frac{1}{\varepsilon} \, F_{1,2}(n) \biggr] \nonumber \\
&& + \biggl[ \ln^2 (-x) \, F_{2,1}(n)  + \frac{1}{\varepsilon} \,\ln (-x) \,
 F_{2,2}(n) + \frac{1}{\varepsilon^2} \, F_{2,3}(n) + \zeta(2)
 \, F_{2,4}(n) \biggr]
\nonumber \\
&& + \biggl[ \ln^3 (-x) \, F_{3,1}(n)  + \frac{1}{\varepsilon} \,\ln^2 (-x) \,
 F_{3,2}(n)  + \frac{1}{\varepsilon^2} \,\ln (-x) \,
 F_{3,3}(n)
+ \frac{1}{\varepsilon^3} \, F_{3,4}(n) \nonumber \\
&&+ \zeta(2)  \,\ln (-x) \, F_{3,5}(n)
+ \zeta(3)  \, F_{3,6}(n) \biggr]
+ \cdots \biggr\},
\label{FI1}
\end{eqnarray}
where $x=q^2/m^2$, $\eta =1$
or $-1$, $c =0$, $1$ and $2$,
and $\alpha=1$ and $2$ for
two-point and three-point cases, respectively.

Here the normalization 
$\hat{N}={(\overline{\mu}^2/m^{2})}^{2\varepsilon}$, 
where $\overline{\mu}=4\pi e^{-\gamma_E} \mu$ is in the standard
$\overline{MS}$-scheme and $\gamma_E$ is the Euler constant.
Moreover, the space-time dimension is $D=4-2\varepsilon$ and
\begin{eqnarray}
C_n ~=~ 1
\label{FI1a}
\end{eqnarray}
for diagrams with one-massive-particle-cuts ($m$-cuts) and
\begin{eqnarray}
C_n  ~=~1, ~~~\mbox{ and }~~~
C_n  ~=~ \frac{(n!)^2}{(2n)!} ~\equiv ~ \hat{C}_n
\label{FI1b}
\end{eqnarray}
for diagrams with
two-massive-particle-cuts ($2m$-cuts).

For  $m$-cut
case,
the coefficients $F_{N,k}(n)$ should have the form
\begin{eqnarray}
F_{N,k}(n) ~ \sim ~ \frac{S_{\pm a,...}}{n^b}\, .
\label{FI1c}
\end{eqnarray}
In this section
$S_{\pm a} \equiv S_{\pm a}(j-1),\ S_{\pm a,\pm b} \equiv
S_{\pm a,\pm b}(j-1),\ S_{\pm a,\pm b,\pm c} \equiv
S_{\pm a,\pm b,\pm c}(j-1)$ are harmonic sums
\begin{eqnarray}
S_{\pm a}(j)\ =\ \sum^j_{m=1} \frac{(\pm 1)^m}{m^a},
\ \ S_{\pm a,\pm b,\pm c,\cdots}(j)~=~ \sum^j_{m=1}
\frac{(\pm 1)^m}{m^a}\, S_{\pm b,\pm c,\cdots}(m),  \label{FI2}
\end{eqnarray}

For  $2m$-cut
case,
the coefficients $F_{N,k}(n)$ should have the form
\footnote{Really, there are even more complicated terms as ones in Eqs.
(58) and (59) of \cite{FleKoVe}, 
which come from other $\eta $ values in (\ref{FI1}).
However,
they are outside of our present consideration.}
\begin{eqnarray}
F_{N,k}(n) ~ \sim ~ \frac{S_{\pm a,...}}{n^b},  ~ \frac{V_{a,...}}{n^b}
,  ~ \frac{W_{a,...}}{n^b}
\label{FI1d}
\end{eqnarray}
where
\begin{eqnarray}
V_{a}(j)\ =\ \sum^j_{m=1}
\, \frac{\hat{C}_m}{m^a},
\ \ V_{a,b,c,\cdots}(j)~=~ \sum^j_{m=1}  \,
\frac{\hat{C}_m}{m^a}\, S_{b,c,\cdots}(m),  \label{FI4} \\
W_{a}(j)\ =\ \sum^j_{m=1} \,
\frac{\hat{C}_m^{-1}}{m^a},
\ \ W_{a,b,c,\cdots}(j)~=~ \sum^j_{m=1}  \,
\frac{\hat{C}_m^{-1}}{m^a}\, S_{b,c,\cdots}(m),  \label{FI5}
\end{eqnarray}

The terms $\sim V_{a,...}$ and $\sim W_{a,...}$
can come only together with the coefficients $C_n =1$ and
$C_n = \hat{C}_n
$, respectively. The terms  $\sim S_{\pm a,...}$
can appear in combination with
both $C_n$ values.
The origin of the appearance of the  terms $\sim V_{a,...}$ and
$\sim W_{a,...}$ in the $2m$-cut
case, is the
product of series (\ref{FI1})
with the different values of the coefficients $C_n =1$ and
$C_n = \hat{C}_n
$.

As
examples, consider two-loop two-point diagrams $I_1$,  $I_5$ and
$I_{12}$, 
studied in \cite{FleKoVe}
\begin{eqnarray}
I_1 &=& \frac{\hat{N}}{q^{2}} \,
\sum_{n=1} \, \frac{x^n}{n} \, \biggl\{
\frac{1}{2} \ln^2 (-x) - \frac{2}{n} \ln (-x)  + \zeta(2)
+2S_2 -2 \frac{S_1}{n} + \frac{3}{n^2} \biggr\} \, ,
\label{FI6a} \\
I_5 &=& \frac{\hat{N}}{q^{2}} \,
\sum_{n=1} \, \frac{(-x)^n}{n} \, \biggl\{
- \ln^2 (-x) + \frac{2}{n} \ln (-x)  -2 \zeta(2)
-4S_{-2} -\frac{2}{n^2} - 2\frac{(-1)^n}{n^2} \biggr\} \, ,
\label{FI6b} \\
I_{12} &=& \frac{\hat{N}}{q^{2}} \,
\sum_{n=1} \, \frac{x^n}{n^2} \, \biggl\{\frac{1}{n} +
 \frac{(n!)^2}{(2n)!} \, \biggl(
-2 \ln (-x)  -3 W_1 + \frac{2}{n} \biggr) \biggr\} \, .
\label{FI6c}
\end{eqnarray}

From (\ref{FI6a}) and (\ref{FI6b}) one can see that the corresponding functions
$F_{N,k}(n)$ have the form
\begin{eqnarray}
F_{N,k}(n) ~ \sim ~ \frac{1}{n^{2-N}},~~~~(N\geq 2),
\label{FI8}
\end{eqnarray}
if we introduce the following complexity of the sums 
($\sum_{i=1}^m a_i =a$)
\begin{eqnarray}
\Phi_{\eta a} \sim \Phi_{\eta a_1, \eta a_2}
\sim \Phi_{\eta a_1,\eta a_{2},\cdots,\eta a_m}
\sim \zeta_{a} \sim \frac{1}{n^a},
\label{FI9}
\end{eqnarray}
where $\Phi=(S,V,W)$.

In Eq. (\ref{FI6c}),
\begin{eqnarray}
F_{N,k}(n) ~ \sim ~ \frac{1}{n^{1-N}},~~~~(N\geq 1),
\label{FI10}
\end{eqnarray}
since now the factor $1/n^2$ has been already extracted.

So, Eqs. (\ref{FI6a})-(\ref{FI6c}) show that the functions $F_{N,k}(n)$
should have the following form
\begin{eqnarray}
\frac{1}{n^{c}} \, F_{N,k}(n) ~ \sim ~ \frac{1}{n^{3-N}},~~~~(N\geq 2)
\label{FI11}
\end{eqnarray}
and the number $3-N$ defines the level of transcendentality (or complexity)
of the coefficients $F_{N,k}(n)$. The property reduces
strongly the number
of the possible elements in $F_{N,k}(n)$.
The level of transcendentality decreases if we consider the
singular parts of diagrams and/or coefficients in front of
$\zeta$-functions and of
logarithm powers.

Other $I$-type integrals in \cite{FleKoVe} have similar form. They have been 
calculated
exactly by differential equation method \cite{Kotikov:1990kg,Kotikov:1991hm}.

Now we consider two-loop three-point diagrams, 
$P_1$, $P_5$, $P_6$, $P_{13}$ and $P_{12}$, considered in \cite{FleKoVe}:
\begin{eqnarray}
P_1 &=& \frac{\hat{N}}{(q^{2})^2} \,
\sum_{n=1} \, \frac{x^n}{n} \, \biggl\{
-\frac{1}{2\ep^3} - \frac{S_1}{\ep^2} + \frac{1}{2\ep}
\biggl[5S_2-S_1^2+ \frac{2}{n^2} - \frac{2}{n} \ln (-x)  
+\frac{1}{2} \ln^2 (-x) - \zeta(2) \biggr]
\nonumber \\ &&
-\frac{8}{3}\zeta_3 -\biggl(S_1+\frac{1}{n}\biggr)\zeta_2
+ \frac{8}{3} S_3 + \frac{9}{2}S_1S_2 + \frac{5}{6}S_1^3 + 4 \frac{S_2}{n}+
2\frac{S_1}{n^2} + \frac{3}{n^3} \nonumber \\
&&+ \biggl(\zeta_2-4S_2-2\frac{S_1}{n}- \frac{3}{n^2}\biggr)\ln (-x)
+ \biggl(S_1 + \frac{3}{2n}\biggr)\ln^2 (-x) -  \frac{1}{2}\ln^3 (-x)
\biggr\} \, ,
\label{FI7a} \\
P_5 &=& \frac{\hat{N}}{(q^{2})^2} \,
\sum_{n=1} \, \frac{(-x)^n}{n} \, \biggl\{
-6\zeta_3 + 2(S_1\zeta_2
+6S_3-2S_1S_2+ 4 \frac{S_2}{n}-\frac{S_1^2}{n}
+ 2\frac{S_1}{n^2} \nonumber \\
&&+ \biggl(-4S_2+S_1^2-2\frac{S_1}{n}\biggr)\ln (-x)
+ S_1\ln^2 (-x)\biggl\} \, ,
\label{FI7b} \\
P_6 &=& \frac{\hat{N}}{(q^{2})^2} \,
\sum_{n=1} \, \frac{(-x)^n}{n} \, \biggl\{
- \frac{1}{\ep^2} \biggl[\ln(-x)-\frac{1}{n}\biggr]
+ \frac{1}{\ep}
\biggl[\zeta_2-3S_2-4S_{-2} - 3 \frac{S_1}{n}
- \frac{3}{n^2}  \nonumber \\
&&+ \biggl(3S_1+ \frac{3}{n}\biggr) \ln (-x)
-\frac{3}{2}\ln^2 (-x) \biggr]
+2\zeta_3 + \biggl(7S_1+\frac{2}{n}\biggr)\zeta_2
-2S_3-9S_1S_2
\nonumber \\&&
+10S_{-3}-12S_{-2,1}-4S_1S_{-2} - \frac{7}{2}\frac{S_2}{n}
-\frac{9}{2}\frac{S_1^2}{n}
-5\frac{S_1}{n^2} -\frac{7}{n^3}
+ \biggl(\frac{7}{2}S_2-\frac{9}{2}S_1^2
\nonumber \\&&
+5\frac{S_1}{n}+\frac{7}{n^2}
-2\zeta_2\biggr)\ln (-x)
+  \frac{1}{2}\biggl(7S_1+\frac{7}{n}\biggr)\ln^2 (-x)
+\frac{7}{6}\ln^3 (-x)\biggl\} \, ,
\label{FI7c} \\
P_{13} &=& \frac{\hat{N}}{(q^{2})^2} \,
\sum_{n=1} \, x^n \,
\biggl\{
-\frac{S_2}{2\ep^2} - \frac{1}{2\ep}
\biggl[S_3+4S_{1,2} -4 \frac{S_2}{n}  \biggr]  +\frac{S_2}{2}\zeta_2
\nonumber \\
&& -S_{1,3}-3S_{3,1}+3S_{1,1,2}+3S_{1,2,1}-S_2^2
+\biggl(7S_3 -8S_{1,2}\biggr) S_1 + \frac{5}{2}S_1^2 S_2
\biggr\} \, ,
\label{FI7d} \\
P_{12} &=& \frac{\hat{N}}{q^{2}} \,
\sum_{n=1} \, \frac{x^n}{n^2} \,
 \frac{(n!)^2}{(2n)!} \, \biggl\{
\frac{2}{\ep^2} + \frac{2}{\ep} \biggl(S_1 -3 W_1 + \frac{1}{n} - \ln (-x)
\biggr) +12 W_2 -18 W_{1,1}
\nonumber \\ && -13S_2 + S_1^2- 6S_1W_1 +2 \frac{S_1}{n} +
\frac{2}{n^2}
-2 \bigg(S_1+\frac{1}{n}\biggr)\ln (-x)  + \ln^2 (-x) \biggr\} \, ,
\label{FI7e}
\end{eqnarray}

 Now the coefficients
$F_{N,k}(n)$ have the form
\begin{eqnarray}
\frac{1}{n^{c}} \, F_{N,k}(n) ~ \sim ~ \frac{1}{n^{4-N}},~~~~(N\geq 3),
\label{FI12}
\end{eqnarray}

The diagrams $P_1$, $P_5$ and $P_6$ (and also $P_3$ in \cite{FleKoVe})
have been calculated
exactly by differential equation method \cite{Kotikov:1990kg,Kotikov:1991hm}.

To find the results for  $P_{13}$ and $P_{12}$ (and also
all others
in \cite{FleKoVe}) we have used the knowledge of the several $n$ terms in the
inverse-mass expansion (\ref{FI1}) (usually less than $n=100$) and
the following arguments (see \cite{FleKoVe1} and discussions therein):
\begin{itemize}
\item
The coefficients should have the structure (\ref{FI12}) with the rule
(\ref{FI9}). The condition (\ref{FI12}) reduces strongly the number of
possible harmonic sums. It should are related with the specific form of the
differential equations for the considered master integrals, like
\bea
\left(\overline{k}\ep  + m^2\frac{d}{dm^2} \right) \, \mbox{ FI } \, = \,
 \mbox{ less complicated diagrams },
\nonumber
\eea
with some $\overline{k}$ values.
We note that for many other master integrals (for example, for sunsets
with two massive lines
in \cite{Kotikov:1990zs,Fleischer:1999hp}) the property 
(\ref{FI12}) is violated: the coefficients
$F_{N,k}(n)$ contain sums with different levels of complexity.
\footnote{Really, Refs. \cite{Kotikov:1990zs,Fleischer:1999hp} contain 
the Nilson polylogarithms,
whose sum of indices relates directly to the level of transcendentality
$(4-N)$. The representation of the series (\ref{FI6a})-(\ref{FI6c}) and
 (\ref{FI7a})-(\ref{FI7e}), containing $S_{\pm a, \cdots}$, as
polylogarithms can be found in \cite{FleKoVe} for $m$-cut case and in 
\cite{Davydychev:2003mv} for
$2m$-cut one, respectively.}
\item
If
a two-loop two-point diagram with the
``similar topology'' (for example, $I_1$ for $P_1$ and $P_3$,  $I_5$ for
$P_5$ and $P_6$, $I_{12}$ for $P_{12}$ an so on)
has been already calculated, we should consider a
similar set of basic elements for corresponding $F_{N,k}(n)$ of
two-loop three-point diagrams
but with the higher level of complexity.
\item
Let the considered diagram contain singularities and/or powers of logarithms.

Because in front of the leading
singularity, or
the largest power of
logarithm, or the largest $\zeta$-function the coefficients are very simple,
they can be often predicted directly from the first several terms of
expansion.

Moreover, often we can calculate the singular part using another technique
(see \cite{FleKoVe} 
for extraction of $\sim W_1(n)$ part). Then we should expand the
singular parts, find the basic elements and try to use them
(with the corresponding increase of the level of complexity) to predict
the regular part of the diagram. If we have to
find the $\ep$-suppressed terms, we should
increase the level of complexity for the corresponding
basic elements.
\end{itemize}

Later, using the ansatz for $F_{N,k}(n)$ and several terms
(usually, less than 100) in the above
expression, which can be calculated exactly, we obtain the system of
algebraic equations for the parameters of the ansatz. Solving the system, we
can obtain the analytical results for FI without
exact calculations.
To check the results, it is needed only to calculate a few more
terms
in the
above inverse-mass expansion (\ref{FI1}) and compare them with the
predictions of our anzatz with the above fixed coefficients.

The arguments give a possibility to find the results for many complicated
two-loop three-point diagrams without direct calculations.
Some variations of the procedure have been successfully used for calculating
the Feynman diagrams for many processes\\ (see 
\cite{Fleischer:1997bw,FleKoVe,Kniehl:2005bc,Kniehl:2006bg}).

\section{Universal anomalous dimension for ${\mathcal N}=4$ SYM}

The final three-loop result
\footnote{
Note, that in an accordance with Ref.~\cite{next}
 our normalization of $\gamma (j)$ contains
the extra factor $-1/2$ in comparison with
the standard normalization (see~\cite{KL})
and differs by sign in comparison with one from Ref.~\cite{VMV}.}
for the universal anomalous dimension $\gamma_{uni}(j)$
for ${\mathcal N}=4$ SYM is~\cite{KLOV}
\begin{eqnarray}
\gamma(j)\equiv\gamma_{uni}(j) ~=~ \hat a \gamma^{(0)}_{uni}(j)+\hat a^2
\gamma^{(1)}_{uni}(j) +\hat a^3 \gamma^{(2)}_{uni}(j) + ... , \qquad \hat a=\frac{\alpha N_c}{4\pi}\,,  \label{uni1}
\end{eqnarray}
where
\begin{eqnarray}
\frac{1}{4} \, \gamma^{(0)}_{uni}(j+2) &=& - S_1,  \label{uni1.1} \\
\frac{1}{8} \, \gamma^{(1)}_{uni}(j+2) &=& \Bigl(S_{3} + \overline S_{-3} \Bigr) -
2\,\overline S_{-2,1} + 2\,S_1\Bigl(S_{2} + \overline S_{-2} \Bigr),  \label{uni1.2} \\
\frac{1}{32} \, \gamma^{(2)}_{uni}(j+2) &=& 2\,\overline S_{-3}\,S_2 -S_5 -
2\,\overline S_{-2}\,S_3 - 3\,\overline S_{-5}  +24\,\overline S_{-2,1,1,1}\nonumber\\
&&\hspace{-1.5cm}+ 6\biggl(\overline S_{-4,1} + \overline S_{-3,2} + \overline S_{-2,3}\biggr)
- 12\biggl(\overline S_{-3,1,1} + \overline S_{-2,1,2} + \overline S_{-2,2,1}\biggr)\nonumber \\
&& \hspace{-1.5cm}  -
\biggl(S_2 + 2\,S_1^2\biggr) \biggl( 3 \,\overline S_{-3} + S_3 - 2\, \overline S_{-2,1}\biggr)
- S_1\biggl(8\,\overline S_{-4} + \overline S_{-2}^2\nonumber \\
&& \hspace{-1.5cm}  + 4\,S_2\,\overline S_{-2} +
2\,S_2^2 + 3\,S_4 - 12\, \overline S_{-3,1} - 10\, \overline S_{-2,2} + 16\, \overline S_{-2,1,1}\biggr)
\label{uni1.5}
\end{eqnarray}
and $S_{a} \equiv S_{a}(j),\ S_{a,b} \equiv S_{a,b}(j),\ S_{a,b,c} \equiv
S_{a,b,c}(j)$ are harmonic sums (see Eq. (\ref{FI2}) and
\begin{eqnarray}
\overline S_{-a,b,c,\cdots}(j) ~=~ (-1)^j \, S_{-a,b,c,...}(j)
+ S_{-a,b,c,\cdots}(\infty) \, \Bigl( 1-(-1)^j \Bigr).  \label{ha3}
\end{eqnarray}

The expression~(\ref{ha3}) is defined for all integer values of arguments
(see~\cite{KK,KL,KoVe})
but can be easily analytically continued to real and complex $j$
by the method of Refs.~\cite{KK,KoVe,AnalCont}.

\subsection{The limit $j\rightarrow 1$ }
The limit $j\rightarrow 1$ is important for the investigation of the small-$x
$ behavior of parton distributions (see review~\cite{Lund} and references
therein). Especially it became popular recently because there are
new experimental data at small $x$ produced by the H1 and ZEUS
collaborations in HERA~\cite{H1}.

Using asymptotic expressions for harmonic sums at $j=1+\omega \rightarrow 1$
(see \cite{KL,KLOV})
we obtain for the ${\mathcal N}=4$ universal anomalous
dimension $\gamma_{uni}(j)$ in Eq.~(\ref{uni1})
\begin{eqnarray}
\gamma _{uni}^{(0)}(1+\omega ) &=&\frac{4}{\omega }+{\mathcal O}\Bigl(\omega^{1}\Bigr),
\label{uni4.1} \\
\gamma _{uni}^{(1)}(1+\omega ) &=&-32\,\zeta _{3}+{\mathcal O}\Bigl(\omega^{1}\Bigr),
\label{uni4.2} \\
\gamma _{uni}^{(2)}(1+\omega ) &=&32\zeta _{3}\,\frac{1}{\omega ^{2}}
-232\,\zeta _{4}\,\frac{1}{\omega }
-1120\zeta _{5}+256\zeta _{3}\zeta _{2}
+{\mathcal O}\Bigl(\omega ^{1}\Bigr)  \label{uni4.3}
\end{eqnarray}
in an agreement with the
predictions for $\gamma _{uni}^{(0)}(1+\omega )$,
$\gamma _{uni}^{(1)}(1+\omega )$ and also for the first term of
$\gamma _{uni}^{(2)}(1+\omega )$ coming from an investigation of BFKL
equation at NLO accuracy in~\cite{KL00}.

\subsection{The limit $j\rightarrow 4$ }

The investigation of the integrability in ${\mathcal N}=4$ SYM for
a BMN-operators~\cite{Berenstein:2002jq} gives a possibility to find the anomalous
dimension of a Konishi operator~\cite{Konishi:1983hf,Beisert:2003tq}, which
has the anomalous dimension
coinciding with our expression~ (\ref{uni1}) for $j=4$
\begin{equation}\label{ADKonTL}
\gamma_{uni}(j)\big|_{j=4}
=-6\, \hat a_s + 24\, \hat a_s^2 - 168\,\hat a_s^3
\end{equation}
It is confirmed also by direct calculation in two~\cite{Arutyunov:2001mh,KoLiVe} and three-loop~\cite{Eden:2004ua}
orders.
The four and five loop corrections to the anomalous
dimension of a Konishi operator have been also calculated recently
in \cite{BaJa,Velizhanin} and \cite{BaHeJaLu,Arutyunov:2010gb}, respectively
(see the recent review \cite{Rej} and references therein).

\subsection{The limit $j\rightarrow  \infty  $ }

In the limit $j\to \infty $ the results~(\ref{uni1.1})-(\ref{uni1.5}) are
simplified significantly. Note, that this limit is related to the study of
the asymptotics of structure functions and cross-sections at $x\rightarrow 1$
corresponding to the quasi-elastic kinematics of the deep-inelastic $ep$
scattering.

We obtain the following
asymptotics for the ${\mathcal N}=4$ universal anomalous dimension $\gamma_{uni}(j)$ in
Eq.~(\ref{uni1}) with
\begin{eqnarray}
\gamma _{uni}^{(0)}(j) &=&-4\Bigl(\ln j+\gamma _{E}\Bigr)+{\mathcal O}\Bigl(j^{-1}
\Bigr),  \label{uni3.1} \\
\gamma _{uni}^{(1)}(j) &=&8\zeta _{2}\,\Bigl(\ln j+\gamma _{E}\Bigr)+12\zeta
_{3}+{\mathcal O}\Bigl(j^{-1}\Bigr),  \label{uni3.2} \\
\gamma _{uni}^{(2)}(j) &=&-88\zeta _{4}\,\Bigl(\ln j+\gamma _{E}
\Bigr)-16\zeta _{2}\zeta _{3}-80\zeta _{5}+{\mathcal O}\Bigl(j^{-1}\Bigr),
\label{uni3.3}
\end{eqnarray}
where $\gamma _{E}$ is Euler constant (see also the normalization
in Eq. (\ref{FI1})).

\subsubsection{Resummation of $\gamma_{uni}$ and the AdS/CFT correspondence}

Last several years
there was a great progress in the investigation of the ${\mathcal N}=4$ SYM
theory in a framework of the AdS/CFT correspondence~\cite{AdS-CFT} where the
strong-coupling limit $\hat{a}_{s}\rightarrow \infty $ is described by
a classical supergravity in the anti-de Sitter space $AdS_{5}\times S^{5}$.
In particular, a very interesting prediction~\cite{15M} (see also~\cite{M})
was obtained for the large-$j$ behavior of the anomalous dimension for
twist-2 operators
\begin{equation}
\gamma (j)=a(z)\,\ln j \,,\qquad\qquad z=\frac{\alpha N_{c}}{\pi } =4\hat a_s
\label{az}
\end{equation}
in the strong coupling regime (see Ref.~\cite{16M} for asymptotic
corrections):
\begin{equation}
\lim_{z\rightarrow \infty }a=-z^{1/2}+\frac{3\ln 2}{8 \pi}+{\mathcal O}
\left(z^{-1/2}\right) \,.
\label{1d}
\end{equation}

On the other hand, 
the results for
$\gamma_{uni}(j)$ in Eqs.~(\ref{uni1}) and (\ref{uni3.1})--(\ref{uni3.3})
allow one to find three first terms of the small-$z$
expansion of the coefficient $a(z)$
\begin{equation}
\lim_{z\rightarrow 0}\,a=-z+\frac{\pi ^2}{12}\, z^2-\frac{11}{720}
\pi^4z^3+...\,.
\end{equation}

For resummation of this series
Lipatov suggested the following equation for the approximation
$\tilde{a}$~\cite{KoLiVe}
\begin{equation}
z=-\widetilde{a}+\frac{\pi ^{2}}{12}\,\widetilde{a}^{2}\,
\label{approx}
\end{equation}
interpolating between its weak-coupling expansion up to NNLO
\begin{equation}
\tilde{a}=-z+\frac{\pi ^{2}}{12}\,z^{2}-\frac{1}{72}\pi ^{4}z^{3}+{\mathcal O}(z^{4})
\end{equation}
and strong-coupling asymptotics
\begin{equation}
\tilde{a} = -\frac{2\sqrt{3}}{\pi} \,z^{1/2} + \frac{6}{\pi^2} + {\mathcal O}
\left(z^{-1/2}\right)
\approx -1.1026\,\,z^{1/2}+0.6079+{\mathcal O}
\left(z^{-1/2}\right).
\end{equation}
It is remarkable, that the predictions for NNLO based on the above simple
equation and obtained before the NNLO results (\ref{uni1.5}) and 
(\ref{uni3.3})), are
valid with the accuracy $\sim 10\%$. It means, that this
extrapolation seems to be good for all values of $z$.
\footnote{Some improvement of (\ref{approx}) van be found in \cite{Smirnov}.}

\subsubsection{Beisert-Eden-Staudacher equation}

Recently 
the integral Beisert-Eden-Staudacher (BES) equation has been proposed
in \cite{BeEdSt}
for some function $f(x)$, which is related with
$a(z)$ of (\ref{az}) at $x=0$, i.e. $f(0)=a(z)$.

At small coupling constant $z$, this equation gives a lot of
coefficients $c_m$ of the expansion
$$
f(0)~= \sum_{m=0} c_m \, \, z^m \, .
$$
These coefficients $c_m$ obey to the 
transcendentality principe,
i.e. $c_m \sim \zeta(2m)$
for $m>0$ (or products of $\zeta$-function with the sum of indices equal to
$2m$). Moreover, up
to 4-loop, the coefficients are in agreement numerically with ones, obtained 
directly from calculations of Feynman diagrams \cite{Smirnov,Bern}.

The most important purpose, however, is to find
the  $z \to \infty $ limit from the BES equation, 
i.e. to try to reproduce 
the Polyakov et al. asymptotics $\sim z^{1/2}$ (see the r.h.s. of
(\ref{1d})). The study was performed and the asymptotics were reproduced
numerically \cite{BeBeKlSc} and analytically \cite{KoLi07}.

Recently authors of
\cite{BaKoKo} found a method to evaluate the
$\tilde{c}_m$ coefficients of the expansion
$$
f(0)~= \sum_{m=0} \tilde{c}_m \, \, z^{(1-m)/2}
$$
of the BES equation and calculated several of them.
The first three coefficients
are in agreement with the results of exact calculations performed
in \cite{15M}, \cite{16M} and \cite{RoTiTs}, respectively.
Moreover, the results of \cite{BaKoKo}  are in well agreement with
transcendentality principe:
$\tilde{c}_1 \sim \ln 2$ and
$\tilde{c}_m \sim \zeta(m)$
for $m>1$ (or products of $\zeta$-function with the sum of indices equal to
$m$).

\section{Bethe-ansatz and four-loop
universal anomalous dimension 
}

The long-range asymptotic Bethe equations for twist-two
operators have the form
\begin{eqnarray}
\label{sl2eq}
\left(\frac{x^+_k}{x^-_k}\right)^2 ~=~
\prod_{m=1,m\neq k}^M \,
\frac{x_k^--x_m^+}{x_k^+-x_m^-}\,
\frac{(1-g^2/x_k^+x_m^-)}{(1-g^2/x_k^-x_m^+)}\,
\exp\left(2\,i\,\theta(u_k,u_j)\right),
\qquad
 \prod_{k=1}^{\hat{M}} \frac{x^+_k}{x^-_k}=1\, .
\end{eqnarray}
These are $\hat{M}$ equations for $k=1,\ldots,\hat{M}$ Bethe roots
$u_k$, which need to be solved for the Bethe roots $u_k$.
The variables $x_k^{\pm}$ are related  to $u_k$ through Zhukovsky map
%
\begin{equation}\label{definition x}
x_k^{\pm}=x(u_k^\pm)\, ,
\qquad
u^{\pm}=u\pm\frac{i}{2}\, ,
\qquad
x(u)=\frac{u}{2}\left(1+\sqrt{1-4\,\frac{g^2}{u^2}}\right),
\end{equation}
The dressing phase $\theta \sim \zeta(3)$
is a rather intricate function conjectured in \cite{BeEdSt} and its exact form
is not so important for the present consideration.

Once the $\hat{M}$ Bethe roots are determined from above equations
for the state of interest, its asymptotic
all-loop anomalous dimension is given by
\begin{equation}
\label{dim}
\gamma^{ABA}(g)=2\, g^2\, \sum^{\hat{M}}_{k=1}
\left(\frac{i}{x^{+}_k}-\frac{i}{x^{-}_k}\right) .
\end{equation}
The above equations
can be solved recursively order by order
in $g$ at arbitrary values of $\hat{M}$ once the one-loop
solution for a given state is known.


This technical problem can nevertheless be surmounted. Assuming the
maximum transcendentality principle \cite{KL} at
four-loop order one may derive the corresponding expression for the
anomalous dimension by making an appropriate ansatz with unknown
coefficients multiplying the nested harmonic sums, and subsequently
fixing these constants. The latter is done by fitting to the exact
anomalous dimensions for a sufficiently large list of specific
values of $\hat{M}$ as calculated from the Bethe ansatz.
\footnote{The study is similar to one considered in the Section 3
and used for calculations of Feynman integrals.}

Luckily, at one-loop the exact solution of the Baxter
equation is known \cite{EdenSt} and is given by a Hahn
polynomial. Knowing the one-loop roots one can then expand equation
(\ref{sl2eq}) in the coupling constant $g$ order by order in
perturbation theory. The equations for the quantum corrections to
the one-loop roots are of course linear, and thus numerically
solvable with high precision.

The result for the four-loop asymptotic dimension has the form \cite{KLRSV}
($\hat{M}=j+2$):
\begin{equation}
\frac{1}{256} \, \gamma^{ABA}_{uni}(j+2) ~=~
\nonumber \\
4\, S_{-7}+6\, S_{7} + ... +
-\zeta(3) S_1(S_3-S_{-3}+2\,S_{-2,1}),
\label{fourloop1}
\end{equation}
where the symbol $...$ marks large set of the nested sums of degree seven.

It is possible 
to analytically continue the expression in the r.h.s. of
 (\ref{fourloop1}) to the vicinity of the pomeron pole at
$M=-1+\omega$.
An explanation for how this is done may be found in
\cite{KoVe}.

Harmonic sums of degree seven
may lead to poles no higher than seventh order in $\omega$.
In fact, it is known that none of the sums in r.h.s. of
(\ref{fourloop1})
can produce such a high-order pole except for the two sums
$S_{7}$ and $S_{-7}$.
Their residues at $1/\omega^7$ are
of opposite sign. Thus, one immediately sees that the sum of the
two residues does {\it not} cancel.

However, from BFKL calculations \cite{KL00,KL}, it is possible to
conclude that at the vicinity of the pomeron pole at
$\hat{M}=-1+\omega$ the four loop anomalous dimensions
\begin{equation}
\gamma_{uni}(1+\omega) ~\sim ~  1/\omega^4 \, .
\label{fourloop4}
\end{equation}

It proves, that the above result is not full and there are so-called wrapping
corrections.

The contribution
of the wrapping
corrections has been added in \cite{BaJaLu}. So, the full result has the
following
form
\begin{eqnarray}
\gamma_{uni}(j+2) &=& \gamma^{ABA}_{uni}(j+2) + \gamma^{wr}_{uni}(j+2),
\nonumber \\
\frac{1}{256} \, \gamma^{wr}_{uni}(j+2) &=&
\frac{1}{2} \, S^2_{1} \, \biggl[ 2\, S_{-5}+2\, S_{5}
+ 4\,\left( \HS_{4,1} - \HS_{3,-2} +
      \HS_{-2,-3} -2\, \HS_{-2,-2,1}\right)
\nonumber \\
& &
- 4\, S_{-2}\zeta(3) -5\, \zeta(5)
\biggr]
\label{wr2}
\nonumber
\end{eqnarray}

This result is in full agreement with BFKL predictions (\ref{fourloop4}).

We note that using similar technique and a property of reciprocity 
(see \cite{Dokshitzer:2006nm} and references therein), the five-loop
corrections fo universal anomalous dimensions have been found in 
\cite{LuReVe}.

\section{Conclusion}
In this review we presented the anomalous dimension
$\gamma _{uni}(j)$ for
the ${\mathcal N}=4$ supersymmetric gauge theory up to the
next-to-next-to-next-to-leading
approximation. All the results have been obtained with using of the
{\it transcendentality principle}.
At the first three orders, the universal anomalous dimension
have been extracted from the corresponding QCD calculations.
The results for four and five loops have been obtained from
the long-range asymptotic Bethe equations
together with some additional terms, so-called {\it wrapping
corrections}, coming in agreement with Luscher approach.\\

A.V.K. thanks to Binur Shaikhatdenov for careful reading of the paper.


\begin{thebibliography}{0}

\bibitem{BFKL}
L.~N.~Lipatov, Sov.\ J.\ Nucl.\ Phys.\ \textbf{23} (1976) 338;
V.~S.~Fadin, E.~A.~Kuraev and L.~N.~Lipatov,
Phys.\ Lett.\ B \textbf{60} (1975) 50;
E.~A.~Kuraev, L.~N.~Lipatov and V.~S.~Fadin,
Sov.\ Phys.\ JETP \textbf{44} (1976) 443;
E.~A.~Kuraev, L.~N.~Lipatov and V.~S.~Fadin,
Sov.\ Phys.\ JETP \textbf{45} (1977) 199;
I.~I.~Balitsky and L.~N.~Lipatov,
Sov.\ J.\ Nucl.\ Phys.\ \textbf{28} (1978) 822;
I.~I.~Balitsky and L.~N.~Lipatov, JETP\ Lett.\ \textbf{30} (1979) 355.


\bibitem{DGLAP}
V.~N.~Gribov and L.~N.~Lipatov, Sov.\ J.\ Nucl.\ Phys.\ \textbf{15} (1972) 438;
V.~N.~Gribov and L.~N.~Lipatov, Sov.\ J.\ Nucl.\ Phys.\ \textbf{15} (1972) 675;
L.~N.~Lipatov, Sov.\ J.\ Nucl.\ Phys.\ \textbf{20} (1975) 94;
G.~Altarelli and G.~Parisi, Nucl.\ Phys.\ \textbf{B126} (1977) 298;
Yu.~L. Dokshitzer, Sov.\ Phys.\ JETP \textbf{46} (1977) 641.

\bibitem{BSSGSO}  L.~Brink, J.~H.~Schwarz and J.~Scherk,
Nucl.\ Phys.\ \textbf{B121} (1977) 77;
F.~Gliozzi, J.~Scherk and D.~I.~Olive,
Nucl.\ Phys.\ \textbf{B122} (1977) 253.


\bibitem{KL}
A.~V.~Kotikov and L.~N.~Lipatov, Nucl.\ Phys.\ \textbf{B661} (2003) 19;
in: {\it Proc. of the XXXV
Winter School}, Repino, S'Peterburg, 2001 (hep-ph/0112346).


\bibitem{VMV}
S.~Moch, J.~A.~M.~Vermaseren and A.~Vogt,
Nucl.\ Phys.\ B {\bf 688} (2004) 101;
A.~Vogt, S.~Moch and J.~A.~M.~Vermaseren,
Nucl.\ Phys.\ B {\bf 691} (2004) 129.


\bibitem{KoLiVe}
A.~V.~Kotikov, L.~N.~Lipatov and V.~N.~Velizhanin,
Phys.\ Lett.\ B \textbf{557} (2003) 114.


\bibitem{next}
V.~S.~Fadin and L.~N.~Lipatov, Phys.\ Lett.\ B \textbf{429} (1998) 127;
G.~Camici and M.~Ciafaloni, Phys.\ Lett.\ B \textbf{430} (1998) 349.

\bibitem{KL00}
A.~V.~Kotikov and L.~N.~Lipatov, Nucl.\ Phys.\ \textbf{B582} (2000) 19.

\bibitem{KLOV}
A.~V.~Kotikov, L.~N.~Lipatov, A.~I.~Onishchenko, and V.~N.~Velizhanin,
Phys.\ Lett.\ B {\bf 595} (2004) 521.

\bibitem{KLRSV}
A.V. Kotikov, L.N. Lipatov, A. Rej, M. Staudacher, and V.N. Velizhanin,
J. Stat. Mech. {\bf 0710} (2007) P10003.

\bibitem{BaJaLu}
Z. Bajnok, R.A. Janik, and T. Lukowski,
Nucl. Phys. B {\bf 816} (2009) 376.

\bibitem{LuReVe}
T. Lukowski, A. Rej, and V.N. Velizhanin,
e-Print: arXiv:0912.1624 [hep-th].

\bibitem{KoReZi}
A.V. Kotikov, A. Rej, and  S. Zieme,
 Nucl. Phys. B {\bf 813} (2009) 460;
M. Beccaria, A.V. Belitsky, A.V. Kotikov, and S. Zieme,
Nucl. Phys. B {\bf 827} (2010) 565.


\bibitem{Staudacher:2004tk}
M.~Staudacher,
JHEP {\bf 0505} (2005) 054;
N. Beisert and M.~Staudacher,
Nucl. Phys. B {\bf 727} (2005) 1.

\bibitem{Beisert}
N. Beisert, Phys. Rept.  {\bf 405} (2005) 1.

\bibitem{Rej}
A. Rej,
J. Phys. A {\bf 42} (2009) 254002.


\bibitem{AdS-CFT}  J.~Maldacena, Adv.\ Theor.\ Math.\ Phys.\ \textbf{2}
(1998) 231; Int.\ J.\ Theor.\ Phys.\ \textbf{38} (1998) 1113;
S.~S.~Gubser, I.~R.~Klebanov and A.~M.~Polyakov, Phys.\
Lett.\ B \textbf{428} (1998) 105;
E.~Witten, Adv.\ Theor.\ Math.\ Phys.\
\textbf{2} (1998) 253.

\bibitem{BFKL2}  A.\ P.\ Bukhvostov, E.\ A.\ Kuraev, L.\ N.\ Lipatov and
G.\ V.\ Frolov,
JETP Lett.\ \textbf{41} (1985) 92;
A.~P.~Bukhvostov, G.~V.~Frolov, L.~N.~Lipatov and E.~A.~Kuraev,
Nucl.\ Phys.\ \textbf{B258} (1985) 601.


\bibitem{N=4}  L.N. Lipatov, Perspectives in Hadronic Physics, in:
\textit{Proc. of the ICTP conf.} (World Scientific, Singapore, 1997).

\bibitem{LN4}
L.N. Lipatov, in: \textit{Proc. of the Int. Workshop on very
high multiplicity physics}, Dubna, 2000, pp.159-176;
L. N. Lipatov, Nucl. Phys. Proc. Suppl. \textbf{99A} (2001) 175.

\bibitem{BDMB} V.~M. ~Braun, S.~E.~Derkachov and A.~N.~Manashov,
Phys.\ Rev.\ Lett. \ \textbf{81} (1998) 2020; 
A.~V.~Belitsky, Phys.\
Lett. \ B \textbf{453} (1999) 59.


\bibitem{Ferretti:2004ba}
G.~Ferretti, R.~Heise and K.~Zarembo,
Phys.\ Rev.\ D {\bf 70} (2004) 074024;
N.~Beisert, G.~Ferretti, R.~Heise and K.~Zarembo,
Nucl. Phys. B {\bf 717} (2005) 137.

\bibitem{Minahan:2002ve}
J.~A.~Minahan and K.~Zarembo,
JHEP {\bf 0303} (2003) 013;
N.~Beisert and M.~Staudacher,
Nucl.\ Phys.\ B {\bf 670} (2003) 439.


\bibitem{Beisert:2003tq}
N.~Beisert, C.~Kristjansen and M.~Staudacher,
Nucl.\ Phys.\ B {\bf 664} (2003) 131.


\bibitem{Integr}  L.~N.~Lipatov, preprint DFPD/93/TH/70;
arXiv:hep-th/9311037, unpublished;
L.~N.~Lipatov,
JETP Lett.\ \textbf{59} (1994) 596 ; 
L.~D.~Faddeev and G.~P.~Korchemsky,
Phys.\ Lett.\ B \textbf{342} (1995) 311.


\bibitem{Fadin:2007xy}
  V.~S.~Fadin and R.~Fiore,
  Phys.\ Lett.\  B {\bf 661} (2008) 139
  [arXiv:0712.3901 [hep-ph]];
  V.~S.~Fadin, R.~Fiore and A.~V.~Grabovsky,
  Nucl.\ Phys.\  B {\bf 831} (2010) 248
  [arXiv:0911.5617 [hep-ph]];
  Nucl.\ Phys.\  B {\bf 820} (2009) 334
  [arXiv:0904.0702 [hep-ph]].


%
\bibitem{LF89}  L.N. Lipatov and V.S. Fadin, Sov. J. Nucl. Phys. {\bf 50}
(1989) 712;
V.S. Fadin, R. Fiore and M.I. Kotsky, Phys. Lett. {\bf B359}
(1995) 181; 
{\bf B387} (1996) 593;
V.S. Fadin and L.N. Lipatov, Nucl. Phys. {\bf B406} (1993)
259;
V.S. Fadin, R. Fiore and A. Quartarolo, Phys. Rev. {\bf D50} (1994) 5893;
V.S. Fadin, R. Fiore and M.I. Kotsky, Phys. Lett. {\bf B389} (1996) 737.


\bibitem{FL96}  V.S. Fadin and L.N. Lipatov, Nucl. Phys. {\bf B477} (1996)
767;
V.S. Fadin, M.I. Kotsky and L.N. Lipatov, Phys. Lett. {\bf B415} (1997) 97.


\bibitem{CCF}  S. Catani, M. Ciafaloni and F. Hautman, Phys. Lett. {\bf
B242} (1990) 97; 
Nucl. Phys. {\bf B366} (1991) 135;
G. Camici and M. Ciafaloni, Phys. Lett. {\bf B386} (1996) 341; 
Nucl. Phys. {\bf B496} (1997) 305;
V.S. Fadin, R. Fiore, A. Flashi and M.I. Kotsky, Phys. Lett. {\bf B422}
(1998) 287.


\bibitem{DRED}  W.~Siegel, Phys.\ Lett.\ B \textbf{84} (1979) 193.



\bibitem{Altarelli}
G. Altarelli, G. Curci, G. Martinelli and S. Petrarca. Nucl. Phys.
B 187 (1981) 461;
 G.A. Shuler, S. Sakakibara and J.G. Korner.
Phys. Lett. B 194 (1987) 125.


\bibitem{FleKoVe}  J. Fleischer, A.V. Kotikov and O.L. Veretin, Nucl. Phys.
\textbf{B547} (1999) 343;





\bibitem{Broadhurst:1987ei}
  D.~J.~Broadhurst,
  Z.\ Phys.\  C {\bf 47} (1990) 115.




\bibitem{Chetyrkin:1981qh}
  K.~G.~Chetyrkin and F.~V.~Tkachov,
  Nucl.\ Phys.\  B {\bf 192} (1981) 159;
  F.~V.~Tkachov,
  Phys.\ Lett.\  B {\bf 100} (1981) 65;
  A.~N.~Vasiliev, Yu.~M.~Pismak and Yu.~R.~Khonkonen,
  Theor.\ Math.\ Phys.\  {\bf 47} (1981) 465
  [Teor.\ Mat.\ Fiz.\  {\bf 47} (1981) 291].




\bibitem{Kotikov:1990zs}
  A.~V.~Kotikov,
  Mod.\ Phys.\ Lett.\  A {\bf 6} (1991) 677.

\bibitem{Kotikov:1990kg}
  A.~V.~Kotikov,
  Phys.\ Lett.\  B {\bf 254} (1991) 158.

\bibitem{Kotikov:1991hm}
  A.~V.~Kotikov,
  Phys.\ Lett.\  B {\bf 259} (1991) 314;
  Phys.\ Lett.\  B {\bf 267} (1991) 123;
  E.~Remiddi,
  Nuovo Cim.\  A {\bf 110} (1997) 1435
  [arXiv:hep-th/9711188].

\bibitem{Kniehl:2005bc}
  B.~A.~Kniehl, A.~V.~Kotikov, A.~Onishchenko and O.~Veretin,
  Nucl.\ Phys.\  B {\bf 738} (2006) 306
  [arXiv:hep-ph/0510235].


\bibitem{Fleischer:1997bw}
  J.~Fleischer, A.~V.~Kotikov and O.~L.~Veretin,
  Phys.\ Lett.\  B {\bf 417} (1998) 163
  [arXiv:hep-ph/9707492].


\bibitem{FleKoVe1}  J. Fleischer, A.V. Kotikov and O.L. Veretin,
Acta Phys. Polon. \textbf{B29} (1998) 2611.


\bibitem{Fleischer:1999hp}
  J.~Fleischer, M.~Y.~Kalmykov and A.~V.~Kotikov,
  Phys.\ Lett.\  B {\bf 462} (1999) 169
  [arXiv:hep-ph/9905249].

\bibitem{Davydychev:2003mv}
  A.~I.~Davydychev and M.~Y.~Kalmykov,
  Nucl.\ Phys.\  B {\bf 699} (2004) 3
  [arXiv:hep-th/0303162].


\bibitem{Kniehl:2006bg}
  B.~A.~Kniehl, A.~V.~Kotikov, A.~I.~Onishchenko and O.~L.~Veretin,
  Phys.\ Rev.\ Lett.\  {\bf 97} (2006) 042001
  [arXiv:hep-ph/0607202];
  A.~Kotikov, J.~H.~Kuhn and O.~Veretin,
  Nucl.\ Phys.\  B {\bf 788} (2008) 47
  [arXiv:hep-ph/0703013];
  B.~A.~Kniehl, A.~V.~Kotikov, Z.~V.~Merebashvili and O.~L.~Veretin,
  Phys.\ Rev.\  D {\bf 79} (2009) 114032
  [arXiv:0905.1649 [hep-ph]];
  B.~A.~Kniehl, A.~V.~Kotikov and O.~L.~Veretin,
  Phys.\ Rev.\ Lett.\  {\bf 101} (2008) 193401
  [arXiv:0806.4927 [hep-ph]];
Phys. Rev. {\bf A80} (2009) 052501 
  [arXiv:0909.1431 [hep-ph]].



\bibitem{KK}  D.~I.~Kazakov and A.~V.~Kotikov, Nucl.\
Phys.\ \textbf{B307} (1988) 721;
[Erratum-ibid.\ \textbf{B345} (1990) 299];
Phys.\ Lett.\ \textbf{B291} (1992) 171.

\bibitem{KoVe}  A.~V.~Kotikov and V.~N.~Velizhanin,
in: {\it Proc. of the XXXIX
Winter School}, Repino, S'Peterburg, 2005 (hep-ph/0501274).

\bibitem{AnalCont}
A.V. Kotikov, Phys. At. Nucl. \textbf{57} (1994) 133.

\bibitem{Lund}  Bo Andersson \emph{et al}., Eur. Phys. J. \textbf{C25}
(2002) 77.

\bibitem{H1}  H1 Collaboration, C. Adloff \emph{et al}., Eur. Phys. J.
\textbf{C21} (2001) 33; 
ZEUS Collaboration, S. Chekanov \emph{et al}., Eur.
Phys. J. \textbf{C21} (2001) 443.

\bibitem{Berenstein:2002jq}
D.~Berenstein, J.~M.~Maldacena and H.~Nastase,
JHEP {\bf 0204} (2002) 013.


\bibitem{Konishi:1983hf}
K.~Konishi,
Phys.\ Lett.\ B \textbf{135} (1984) 439.

\bibitem{Arutyunov:2001mh}
G.~Arutyunov, B.~Eden, A.C.~Petkou and E.~Sokatchev,
Nucl.\ Phys.\ B {\bf 620} (2002) 380;
B.~Eden, C.~Jarczak, E.~Sokatchev and Y.~S.~Stanev,
Nucl. Phys. B {\bf 722} (2005) 119;
B.~Eden,
Nucl. Phys. B {\bf 738} (2006) 409.


\bibitem{Eden:2004ua}
B.~Eden, C.~Jarczak and E.~Sokatchev,
Nucl. Phys. B {\bf 712} (2005) 157.


\bibitem{BaJa}
Z. Bajnok and R.A. Janik,
Nucl. Phys. B {\bf 807} (2009) 625.


\bibitem{Velizhanin}
F. Fiamberti, A. Santambrogio, C. Sieg, and D. Zanon,
Phys. Lett. B {\bf 666} (2008) 100;
V.N. Velizhanin, JETP Lett. {\bf 89} (2009) 6;
e-Print: arXiv:0808.3832 [hep-th].

\bibitem{BaHeJaLu}
Z. Bajnok, A. Hegedus, R.A. Janik, and T. Lukowski,
Nucl. Phys. B {\bf 827} (2010) 426.


\bibitem{Arutyunov:2010gb}
  G.~Arutyunov, S.~Frolov and R.~Suzuki,
  arXiv:1002.1711 [hep-th];
  J.~Balog and A.~Hegedus,
  arXiv:1002.4142 [hep-th].

\bibitem{15M}  S.\thinspace S.~Gubser, I.\thinspace R.~Klebanov and
A.\thinspace M.~Polyakov,
Nucl.\ Phys.\ \textbf{B636} (2002) 99.


\bibitem{M}
M.~Kruczenski,
JHEP \textbf{0212} (2002) 024;
Yu.~Makeenko,
JHEP \textbf{0301} (2003) 007
[arXiv:hep-th/0210256];
M.~Axenides, E.~Floratos and A.~Kehagias,
Nucl.\ Phys.\ \textbf{B662} (2003) 170
[arXiv:hep-th/0210091].


\bibitem{16M}  S.~Frolov and A.\thinspace A.~Tseytlin, JHEP \textbf{0206}
 (2002) 007.

\bibitem{BeEdSt}
N. Beisert, B. Eden, and  M. Staudacher,
J. Stat. Mech. {\bf 0701} (2007) P021.

\bibitem{Smirnov} Z. Bern, M. Czakon, L. Dixon, D. Kosover and V. Smirnov,
 Phys. Rev. \ D {\bf 75} (2007) 085010.

\bibitem{Bern}
Z. Bern, J.J.M. Carrasco, H. Johansson and  D.A. Kosower,
Phys. Rev. \ D {\bf 76} (2007) 125020


\bibitem{BeBeKlSc}
M.K. Benna, S. Benvenuti, I.R. Klebanov, and A. Scardicchio,
Phys. Rev. Lett. {\bf 98} (2007) 131603.

\bibitem{KoLi07}
A.V. Kotikov and L.N. Lipatov,
Nucl. Phys. B {\bf 769} (2007) 217.

\bibitem{BaKoKo}
B. Basso, G.P. Korchemsky, and J. Kotanski,
Phys. Rev. Lett. {\bf 100} (2008) 091601.


\bibitem{RoTiTs}
R. Roiban, A. Tirziu, and  A.A. Tseytlin,
JHEP {\bf 0707} (2007) 056.


\bibitem{EdenSt}
B.~Eden and M.~Staudacher,
J. Stat. Mech. {\bf 0611} (2006) P014.


\bibitem{Dokshitzer:2006nm}
  Yu.~L.~Dokshitzer and G.~Marchesini,
  Phys.\ Lett.\  B {\bf 646} (2007) 189
  [arXiv:hep-th/0612248];
  B.~Basso and G.~P.~Korchemsky,
  Nucl.\ Phys.\  B {\bf 775} (2007) 1
  [arXiv:hep-th/0612247];
  M.~Beccaria,
  JHEP {\bf 0709} (2007) 023
  [arXiv:0707.1574 [hep-th]];
  M.~Beccaria and V.~Forini,
  JHEP {\bf 0903} (2009) 111
  [arXiv:0901.1256 [hep-th]];
  M.~Beccaria, V.~Forini, T.~Lukowski and S.~Zieme,
  JHEP {\bf 0903} (2009) 129
  [arXiv:0901.4864 [hep-th]];
  M.~Beccaria, V.~Forini and G.~Macorini,
  arXiv:1002.2363 [hep-th].


\end{thebibliography}
\end{document}